\documentclass[%
  aip,
  amsmath,amssymb,
  reprint,%
  nofootinbib,
  floatfix,
]{revtex4-1}

\usepackage{graphicx}
\usepackage{dcolumn}
\usepackage{bm}
\usepackage{subfigure}
\usepackage{xcolor}
\usepackage{comment}
\usepackage[normalem]{ulem}

\usepackage[utf8]{inputenc}
\usepackage[T1]{fontenc}
\usepackage{mathptmx}
\usepackage[colorlinks=true, allcolors=blue]{hyperref}
\usepackage{url}
\usepackage{lineno}
\usepackage{wasysym} 

\newcommand\orcid[1]{\href{https://orcid.org/#1}{\includegraphics[height=0.8em]{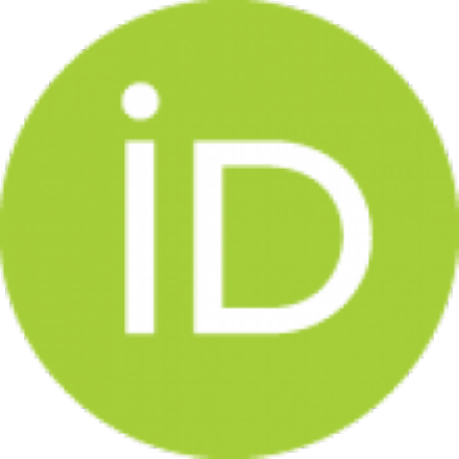}}}

\newcommand{\nocontentsline}[3]{}
\newcommand{\tocless}[2]{\bgroup\let\addcontentsline=\nocontentsline#1{#2}\egroup}

\begin{document}

\title{The Simons Observatory: Large Diameter and Large Load-Capacity Superconducting Magnetic Bearing for a Millimeter-Wave Polarization Modulator}

\author{Daichi\,Sasaki\,\orcid{0009-0003-2513-2608}}
\email[Authors to whom correspondence should be addressed:
\href{mailto: daichi.sasaki@phys.s.u-tokyo.ac.jp}{daichi.sasaki@phys.s.u-tokyo.ac.jp}]{}
\affiliation{Department of Physics, Graduate School of Science, The University of Tokyo, Tokyo 113-0033, Japan\looseness=-1}
\author{Junna\,Sugiyama\,\orcid{0009-0007-7435-9082}}
\affiliation{Department of Physics, Graduate School of Science, The University of Tokyo, Tokyo 113-0033, Japan\looseness=-1}
\author{Kyohei\,Yamada\,\orcid{0000-0003-0221-2130}}
\affiliation{Joseph Henry Laboratories of Physics, Jadwin Hall, Princeton University, Princeton, NJ 08544, USA\looseness=-1}
\author{Bryce\,Bixler\,\orcid{0009-0008-4312-6814}}
\affiliation{Department of Physics, University of California, San Diego, La Jolla, CA 92093, USA}
\author{Yuki\,Sakurai\,\orcid{0000-0001-6389-0117}}
\affiliation{Department of Mechanical and Electrical Engineering, The Suwa University of Science, Nagano 391-0213, Japan\looseness=-1}
\affiliation{Kavli Institute for the Physics and Mathematics of the Universe (WPI), UTIAS, The University of Tokyo, Chiba 277-8583, Japan\looseness=-1}
\author{Kam\,Arnold\,\orcid{0000-0002-3407-5305}}
\affiliation{Department of Physics, University of California, San Diego, La Jolla, CA 92093, USA}
\author{Bradley\,R.\,Johnson\,\orcid{0000-0002-6898-8938}}
\affiliation{Department of Astronomy, University of Virginia, Charlottesville, VA 22904, USA}
\author{Akito\,Kusaka\,\orcid{0009-0004-9631-2451}}
\affiliation{Department of Physics, Graduate School of Science, The University of Tokyo, Tokyo 113-0033, Japan\looseness=-1}
\affiliation{Physics Division, Lawrence Berkeley National Laboratory, Berkeley, CA 94720, USA}
\affiliation{Kavli Institute for the Physics and Mathematics of the Universe (WPI), UTIAS, The University of Tokyo, Chiba 277-8583, Japan\looseness=-1}
\affiliation{Research Center for the Early Universe, School of Science, The University of Tokyo, Tokyo 113-0033, Japan\looseness=-1}

\date{\today}

\begin{abstract}

We present the design methodology and characterization of a superconducting magnetic bearing (SMB) system for the polarization modulator in the SAT-LF, one of the small aperture telescopes (SATs) in the Simons Observatory (SO) that is sensitive at 30/40\,GHz frequency bands.
SO is a ground-based cosmic microwave background (CMB) polarization experiment, with the SATs specifically aiming to search for primordial parity-odd polarization anisotropies at degree scales.
Each SAT is equipped with a cryogenic, continuously rotating half-wave plate (HWP) as a polarization modulator to mitigate atmospheric $1/f$ noise and instrumental systematics.
The HWP system employs an SMB, consisting of a ring-shaped magnet and superconductor, to achieve a 550\,mm clear aperture and stable 2\,Hz rotation at a temperature of around 50\,K.
One challenge for the HWP system in the SAT-LF is the large 35\,kg load on the SMB due to the thicker HWP than in previous telescopes.
Since the SMB stiffness is critical for maintaining the alignment of the HWP in the telescope, we developed a method to quantitatively predict the stiffness using finite element simulations with the so-called H-formulation.
We evaluated the stiffness for various geometries of the magnet and superconductor, thereby optimizing their dimensions.
The prediction is in excellent agreement with experimental measurements of the fabricated SMB, demonstrating a $\sim$5\% accuracy.
We also demonstrated that the SMB achieves sufficiently low friction-induced heat dissipation, measured at 0.26\,W when rotating at 2\,Hz.
The design methodology and the implementation of the SMB demonstrated here not only provides an enabling technology for the SO SAT-LF, but also is a crucial stepping stone for future CMB experiments that make use of HWP polarization modulators.

\end{abstract}

\maketitle


\section{Introduction}
\label{sec:intro}

\begin{figure*}[t]
    \begin{center}
    \includegraphics[width = 0.85\textwidth]{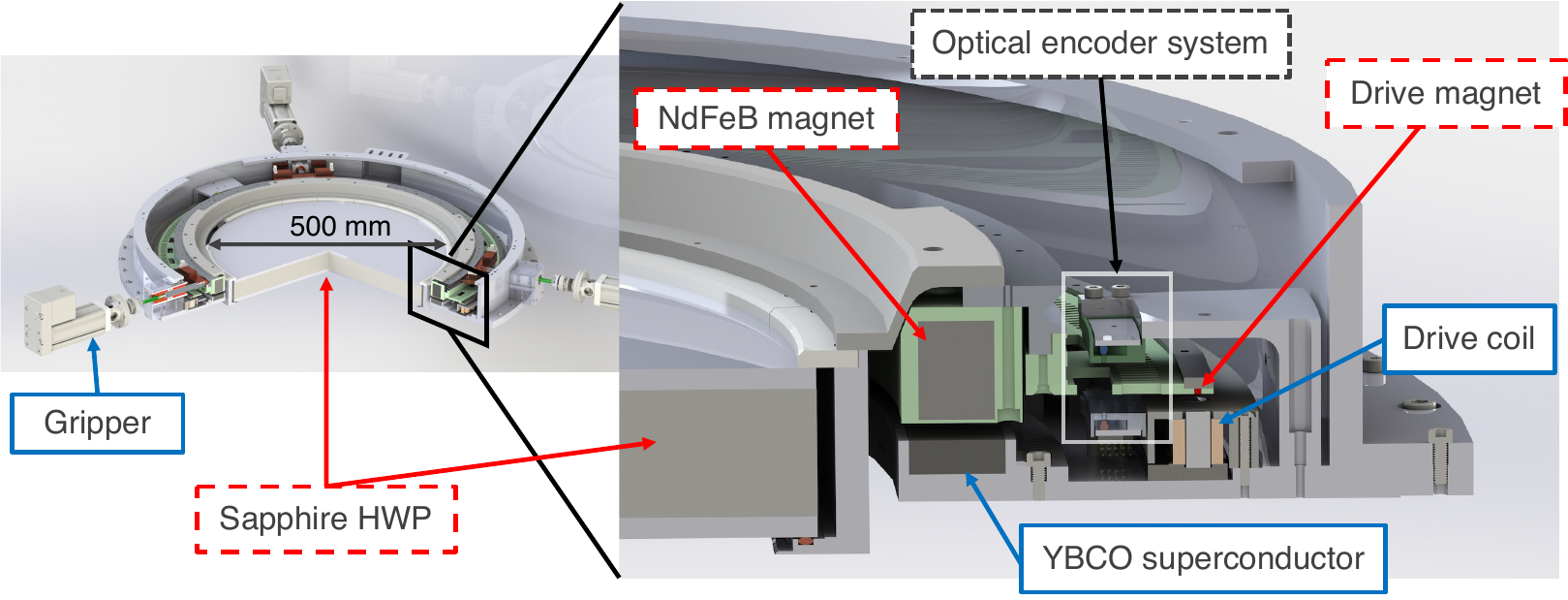}
    \caption{Left panel: CAD views of the SAT polarization modulator. Right panel: Magnified cross-sectional view of the SMB and rotation mechanism. Rotating components are labeled with red dashed boxes, while stationary components are labeled with blue boxes. The optical encoder system is configured from top to bottom with photodiodes, an encoder plate with slits, and LEDs.}
    \label{fig:design}
    \end{center}
\end{figure*}

The cosmic microwave background (CMB) is the oldest electromagnetic radiation observable in the universe, originating from the epoch of recombination.
Observations of its intensity and polarization anisotropies have revealed a wealth of cosmological information. 
However, despite being one of the most intriguing targets in cosmology, the possible parity-odd polarization patterns generated by primordial tensor perturbations\cite{kamionkowski_probe_1997, seljak_signature_1997}, referred to as primordial B-modes, remain undetected.
Primordial tensor perturbations could have been produced during an exponential expansion phase in the early universe, known as inflation. \cite{abbott1986inflationary, linde2005particle, Linde2007}
A measurement of the primordial B-mode would provide a unique opportunity to constrain models of the early universe and offer insights into physics at the grand unified theory scale.\cite{seljak_direct_1999}

The Simons Observatory (SO) is a ground-based CMB experiment located in the Atacama Desert in Chile.\cite{Ade2019}
Its small aperture telescopes (SATs) are an array of millimeter-wave telescopes with aperture sizes of 420 mm.\cite{Ali2020, Kiuchi2020, Galitzki_2024}
The SATs specifically aim to measure or constrain the amplitude of primordial B-modes by observing CMB polarization at degree scales.

Observing in a wide frequency range is necessary for separating CMB signal from galactic foregrounds.
To date, SO has deployed three SATs: two for the mid-frequency bands (SAT-MF, 90/150 GHz) and one for the ultra-high frequency bands (SAT-UHF, 220/280 GHz).
In addition to these, we are currently developing a new SAT that will observe the low-frequency band (SAT-LF, 30/40 GHz).
The polarization signals in this frequency band are dominated by galactic synchrotron emission.
It is essential to observe this emission with unprecedented sensitivity using the SAT-LF so that we can remove foregrounds contamination from the primordial B-mode signal.

Modulating the polarization signal is one of the key techniques for precise polarization measurements.
It suppresses atmospheric $1/f$ noise and mitigates instrumental systematic effects.
Various polarization modulation methods have been used in past CMB experiments.\cite{Map_radiometer_2003, POLAR_2003, DASI_2005, Barkats_2005, matsumura_phd, johnson_maxipol_2007, Chen_2009, klein_cryogenic_2011, planck_lf_instrument, QUIET_2012, Moyerman_2013, kusaka_modulation_2014, Bryan_2016, Miller_2016, hill_design_2016, Bryan_2016, johnson_large-diameter_2017, GB2020, HillPB2bCHWP2020, Harrington_2021}

The SATs employ a cryogenic, continuously rotating half-wave plate (HWP) as the polarization modulator.\cite{Yamada2024-ro, sugiyama2024simons}
The HWP, made of a birefringent material, flips the incident polarization relative to a specific axis fixed to the plate.
Polarization modulation is achieved by continuously rotating the HWP at a few hertz.
In each SAT, we use a broadband HWP achieved by stacking three sapphire plates.\cite{pancharatnam_achromatic_1955,sugiyama2024simons}
The HWP must have a diameter of 500\,mm and be maintained below 85\,K to suppress its thermal emission, which otherwise would raise the photon noise on the detectors.
Therefore, the SATs require a rotation mechanism that can operate at cryogenic temperatures with a large clear aperture.
We employ a superconducting magnetic bearing (SMB) to achieve this capability.

An SMB is a contactless bearing that exploits the pinning force between a type-II superconductor and a magnet.\cite{Hull_2000}
It can operate at cryogenic temperatures under vacuum and offers several advantages over conventional mechanical bearings, including reduced friction, lower vibration, and long-term durability with minimal maintenance.
SMBs have been used for various applications, such as flywheel energy storage\cite{HULL1994449}, high-efficiency motor bearings, and polarization modulators in previous CMB experiments.\cite{hanany2003cosmic, Hull_2005, klein2011cryogenic, HillPB2bCHWP2020, Sakurai_2020}
So far, SO has deployed three SMBs, two in the SAT-MFs and one in the SAT-UHF.\cite{Yamada2024-ro}

We also employ an SMB in the SAT-LF.
One of the key challenges in its development is the increased thickness of the HWP for the SAT-LF due to the larger observational wavelength, measuring around 10\,mm.
The resultant HWP stack has a total thickness exceeding 30\,mm, making it significantly heavier than those in the SAT-MFs and SAT-UHF.
As a result, the SMB must support a load of about 35\,kg, excluding the weight of the bearing assembly itself.
This increased load introduces a challenge in maintaining the alignment of the rotor within acceptable limits during telescope operations.

There is no precedent for such demanding challenges posed by the large aperture and large load in the development of cryogenic polarization modulators at microwave frequencies.
Moreover, there is no simple scaling law to predict how the bearing force changes with magnetic field strength or superconductor size.
In fields where SMBs are utilized, finite element simulations are commonly used in design process to meet specific requirements.\cite{YILDIZ201964, LI2024103849}
In this study, we introduce, for the first time, quantitative predictions of rotor alignment based on finite element simulations into the development of a cryogenic HWP system and establish a design methodology based on these predictions.
We report on the development and characterization of the SMB for the SAT-LF polarization modulator, which was realized using this approach.

Section\,\ref{sec:des_req} provides an overview of the design and the bearing requirements.
Section\,\ref{sec:sim_opt} describes the simulation method and how the SMB geometry is optimized.
Section\,\ref{sec:impl_char} details the SMB implementation and performance measurements.
Section\,\ref{sec:con} summarizes the development.

\section{System Overview}
\label{sec:des_req}

\subsection{Overview of the SAT Polarization Modulator}
\label{subsec:design}

Figure\,\ref{fig:design} shows the design of the SAT-LF polarization modulator.
In the SAT-LF, we modified the dimensions of the magnet and the superconductor, but overall rotation mechanism remains the same as described in \citet{Yamada2024-ro}.

The main components of the SMB are the ring-shaped neodymium (NdFeB; N52) magnet assembly and the ring-shaped yttrium barium copper oxide (YBCO) superconductor assembly.
The HWP is attached to the magnet ring assembly, and together with the encoder ring, they constitute the rotor.
The magnet ring is composed of arc-shaped segments magnetized in a direction parallel to the rotation axis.
The YBCO ring assembly has the same diameter as the magnet ring and is placed beneath it.
During cooldown, the rotor is held in place by a grip and release mechanism, hereafter gripper, so the YBCO is field-cooled in the magnet's field.
Once the YBCO enters its superconducting state, the rotor is released and levitates via the pinning force between the magnet and YBCO.
This pinning force constrains the rotor's movement except for its rotation around the optical axis, which is the symmetry of the magnetic field.

The rotation of the rotor is driven by a linear motor.
The stator has 120 coils, that produce a magnetic field phase-shifted by 120 degrees relative to one aother, synchronized with the rotor's rotation.
Eighty drive magnets, attached to the encoder ring with alternating polarity, experience torque from this times-varying magnetic field.
The encoder ring also has 569 slits around its circumference.
Infrared LEDs and photodiodes are used to measure the rotor angle by detecting light chopped by these slits.
Two such optical encoder subassemblies are installed at positions separated by 180\,degrees.

\subsection{Requirements}
\label{subsec:req}

One of the key requirements for the SMB is maintaining alignment during telescope motion.
The SAT is mounted on a three-axis platform, allowing rotation in azimuth, elevation, and around the line of sight (boresight rotation). 
The telescope moves between elevation angles of 50 and 90\,degrees and may perform boresight rotation during nominal operation.
Since the  SMB stiffness is finite, lateral rotor displacement occurs as the telescope tilts.
We require the rotor displacement to remain under 4\,mm; larger displacements could degrade drive motor efficiency, or lead to physical contact or interference in the optical path.\footnote{We assume an initial alignment accuracy under 1\,mm when held by the gripper, resulting in a total displacement tolerance of 5\,mm.}
Accordingly, the SMB's lateral stiffness must be increased over those in the SAT-MFs and SAT-UHF due to the heavier HWP in the SAT-LF.

Another requirement is that the rotor must rotate at a minimum of 2\,Hz to effectively mitigate atmospheric noise.
Magnetic field uniformity is crucial for two reasons: reducing friction during rotation and facilitating the initiation of rotation.
Because the total heat dissipation of the SMB system is limited to 3\,W given the cooling capacity of the cryostat, frictional heat dissipation from the rotating SMB at 2\,Hz must remain within acceptable limits.
SMB friction arises from hysteresis loss in the superconductor and eddy current loss in surrounding conductors\cite{Hull_2000}, both of which can be reduced by improving magnetic field uniformity.
Furthermore, improved uniformity lowers the potential energy barrier when the rotor is static, thus facilitating startup rotation with less current supplied to the drive coils.
A specific requirement is to start the rotation with the total current being less than 1\,A.

We address the alignment requirement by quantitatively predicting the SMB stiffness using finite element simulations.
The rotational issue is handled in combination with the alignment issue as follows: 
First, simulations allow us to optimize the geometries of the magnet and superconductor to increase stiffness.
This enables a larger distance between the magnet and the superconductor while maintaining sufficient stiffness.
By combining this with our analytical model of magnetic field non-uniformity, we find that the field non-uniformity at the superconductor's position improves, which is advantageous for meeting the rotational requirement.
Therefore, our fundamental strategy is to optimize the geometries of the magnet and superconductor to enhance stiffness.

\section{Design}
\label{sec:sim_opt}

\subsection{Bearing Stiffness Simulation}
\label{subsec:sim}

\begin{figure}
    \centering
    \includegraphics[width = 0.36\textwidth]{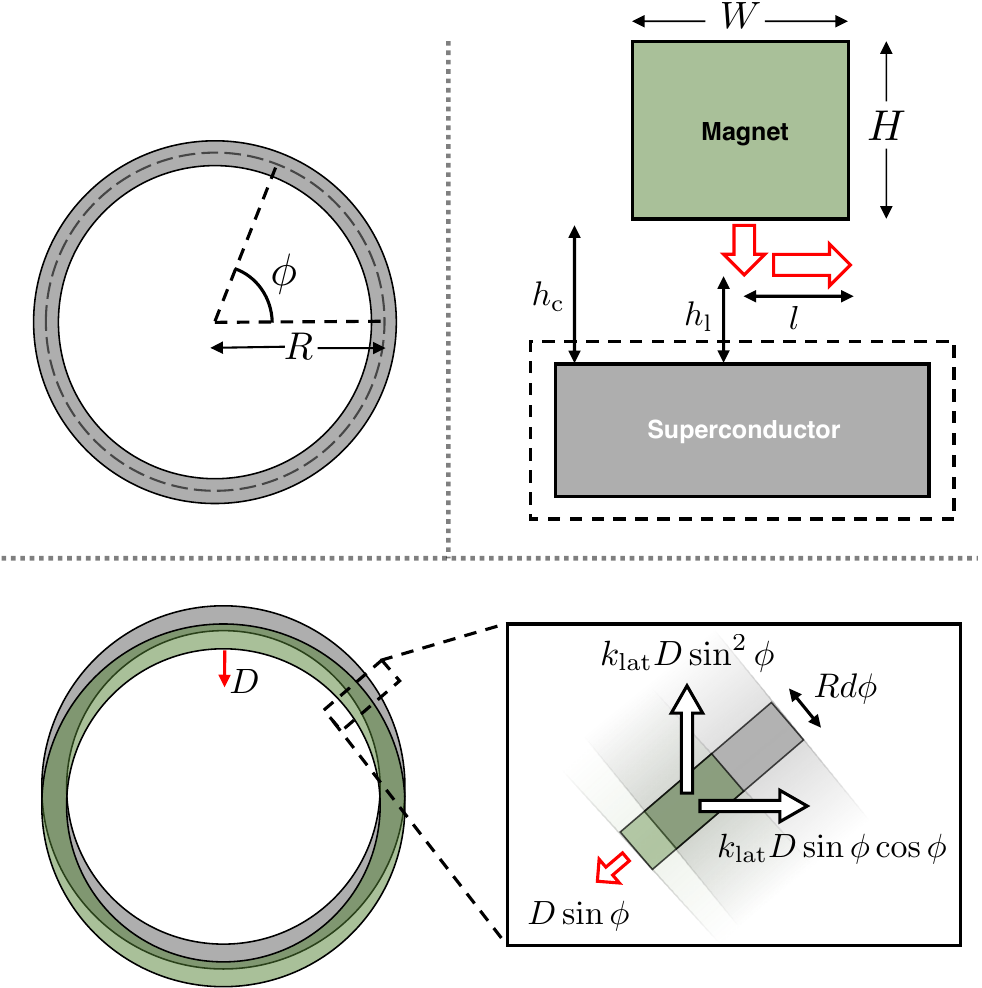}
    \caption{Top left panel: Diagram of the YBCO ring viewed from above, showing the definitions of $R$ and $\phi$. Top right panel: Conceptual diagram of the simulation using the H-formulation. The dashed square is the cross-section of the YBCO and its surroundings, representing the simulation domain. The magnet is not included in the simulation domain, but its motion (red arrows) is emulated through time-dependent boundary conditions along the dashed lines. The distance between the magnet and the superconductor is $h_{\mathrm{c}}$ at the time of the superconducting transition. It becomes $h_{\mathrm{l}}$ after the rotor is released, balancing the where the gravitational force. The lateral displacement is $l$. Bottom panel: Conceptual diagram of the lateral stiffness calculation. The green ring represents the magnet. The enlarged diagram depicts a small element of the YBCO and the magnet, taken around a certain angle $\phi$, corresponding to the top-down view of the situation shown in the top-right panel. When the overall ring displacement is $D$, each segment's displacement is $D\sin\phi$, and the force in the global displacement direction is $k_{\mathrm{lat}}D\sin^2\phi R\,d\phi$. The term perpendicular to it is integrated to zero.}
    \label{fig:sim}
\end{figure}

Here, we describe the method of finite element simulation and how to derive the force, stiffness, and alignment of the SMB based on its results.

For the finite element simulation, we adopted the 2D version of the method called H-formulation, which is detailed in \citet{Queval_2018} and \citet{Sass_2015}.
We employed COMSOL Multiphysics® and used the same parameters for YBCO as those specified for the 2D axisymmetric and 3D cases in \citet{Queval_2018}.

As shown in the top left panel of Fig.\,\ref{fig:sim}, let $R$ be the radius of the YBCO ring, and $\phi$ be its circumferential coordinate.
We define $h$ as the distance between the magnet and the YBCO surfaces.
Specifically, we define the cooling height, $h_\mathrm{c}$, and the levitation height, $h_{\mathrm{l}}$, as follows:
$h_\mathrm{c}$ is the value of $h$ when the rotor is fixed by the gripper and the YBCO transitions into the superconducting state.
After the rotor is released to levitate, $h$ decreases to $h_{\mathrm{l}}$, determined by the balance between the SMB kevitation force and the gravitational force acting on the rotor.
We denote the magnet cross-sectional width and height as $W$ and $H$, respectively, and consider a lateral displacement $D$ in the $\phi = -\pi/2$ direction.

Instead of performing a full 3D simulation of the entire system, we approximate it by 2D simulations.
This approximation assumes that the force acting on an infinitesimal arc segment of the YBCO and magnet is equivalent to the force per infinitesimal length of the YBCO and magnet when they are modeled as infinitely long rods with the same cross-section as the ring.
Since the simulation of infinitely long rods can be conducted in 2D, this offers advantages in terms of computational cost.
This approximation is valid because $R\sim 290\,\mathrm{mm}$ is large compared to $W \lesssim 30\,\mathrm{mm}$ and $D\lesssim 5\,\mathrm{mm}$.

In the H-formulation, time-dependent partial differential equations of the magnetic field are solved in the YBCO and its surrounding thin region.
The magnet'field is simulated separately in advance, and its motion is emulated by applying time-dependent magnetic field boundary conditions on the dashed lines in the top right panel of Fig.\,\ref{fig:sim}.
Within a type-II superconductor, the magnetic flux quanta are fixed in a low-energy configuration, causing the macroscopic magnetic field to remain at its value during the transition.
Therefore, when the external magnetic field changes due to the motion of the magnet, currents are induced in the superconductor.
The force between the superconductor and the magnet is then calculated as the electromagnetic force exerted by the magnetic field on these induced currents.

We separate the simulation into two parts. 
First, we compute the levitation force $F_{\mathrm{lev}}$ as a function of the magnet-YBCO distance $h$.
In this step, the magnet is moved from the cooling height $h_{\mathrm{c}}$ towards the YBCO, and the simulation yields the vertical force per unit length of the rods, $f_{\mathrm{lev}}(h)$.
Since $f_{\mathrm{lev}}(h)R\,d\phi$ corresponds to the force for the infinitesimal element with angle $d\phi$, the total force is
\begin{equation}
F_{\mathrm{lev}}(h) = \int_0^{2\pi}f_{\mathrm{lev}}(h) R\,d\phi = 2\pi R f_{\mathrm{lev}}(h).
\end{equation}
We find $h_{\mathrm{l}}$ by solving the equilibrium condition
\begin{equation}
F_{\mathrm{lev}}(h_{\mathrm{l}}) = mg\sin\theta, \label{eq:Flev}
\end{equation}
where $m$ is the rotor mass, $g$ is the acceleration of gravity, and $\theta$ is the elevation angle of the telescope.
The vertical stiffness is defined as $K_{\mathrm{lev}} = dF_{\mathrm{lev}}/dh|_{h=h_{\mathrm{l}}}$.

Once $h_{\mathrm{l}}$ is determined, we proceed to the second simulation that calculates the lateral force.
In this simulation, the magnet is first moved vertically from $h_{\mathrm{c}}$ to $h_{\mathrm{l}}$, after which it is moved laterally.
The lateral force per unit length, $f_{\mathrm{lat}}$, is calculated as a function of the lateral displacement $l$.
Within the displacement range of interest ($<$5\,mm), $f_{\mathrm{lat}}$ is proportional to $l$.
Thus, we calculate the lateral stiffness per unit length, $k_{\mathrm{lat}}$, using the relation $f_{\mathrm{lat}}=k_{\mathrm{lat}}l$.
For displacements of interest ($<$5\,mm), $f_{\mathrm{lat}}$ is proportional to $l$, and we calculate the lateral stiffness per unit length, $k_{\mathrm{lat}}$, using the relation $f_{\mathrm{lat}} = k_{\mathrm{lat}}l$.
The total lateral force for a ring displacement $D$ is 
\begin{equation}\label{eq:Flat}
F_{\mathrm{lat}} = \int_0^{2\pi}k_{\mathrm{lat}}D\sin^2\phi R\,d\phi=\pi R k_{\mathrm{lat}}D.
\end{equation}
As shown in the bottom panel of Fig.\,\ref{fig:sim}, the term $\sin^2\phi$ arises because the lateral displacement for each infinitesimal element is $D\sin\phi$ and another $\sin\phi$ factor arises when extracting the force in the direction of the overall displacement.
The total lateral stiffness is defined as $K_{\mathrm{lat}} = dF_{\mathrm{lat}}/dD = \pi R k_{\mathrm{lat}}$.
At telescope elevation $\theta$, the rotor displacement $D(\theta)$ becomes
\begin{equation}\label{eq:displacement}
D(\theta) = \frac{mg\cos\theta}{K_{\mathrm{lat}}}.
\end{equation}

Here, since $K_{\mathrm{lat}}$ depends on $\theta$ through the \(\theta\) dependence of \(h_{\mathrm{l}}\) in Eq.\,\eqref{eq:Flev}, the $\theta$ dependence of $D(\theta)$ is not straightforward.
However, within the range of interest (50 to 90\,degrees), the fractional difference in $K_{\mathrm{lat}}$ is approximately 2\%. (see Appendix\,\ref{sec:klat})
This allows us to treat $K_{\mathrm{lat}}$ as constant when evaluating it from the displacement measurements at different elevation angles in Sec.\,\ref{subsec:stiff_meas}.

\subsection{Optimally Stiff Magnet Geometry}
\label{subsec:opt}

\begin{figure}
    \centering
    \includegraphics[width = 0.40\textwidth]{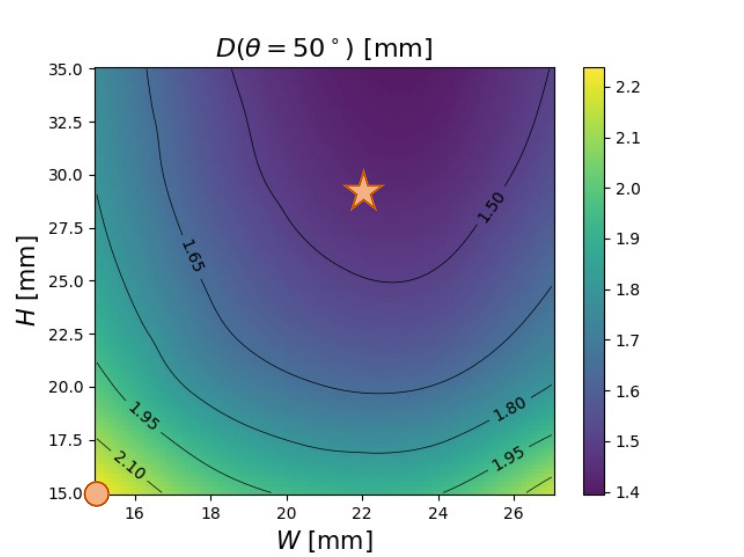}
    \caption{Simulated lateral displacement at an elevation of 50\, degrees for different magnet cross-sectional dimensions, with the YBCO cross-section fixed at 28\,mm$\times$10\,mm and $h_\mathrm{c}=6\,\mathrm{mm}$. The horizontal axis is the magnet width ($W$), and the vertical axis is the magnet height ($H$). The circle indicates the geometry used in the SAT-MFs and SAT-UHF (15\,mm $\times$ 15\,mm), and the star mark indicates the newly chosen geometry (22\,mm $\times$ 29\,mm).}
    \label{fig:opt}
\end{figure}

We performed a series of simulations to optimize magnet and superconductor dimensions to improve stiffness.
Specifically, for each geometry, we compute the rotor's lateral displacement at $\theta = 50^\circ$ by applying Eq.\,\eqref{eq:displacement} with the $K_{\mathrm{lat}}$ from the finite element simulations.
We assume the rotor mass is $m = 35\,\mathrm{kg}$, excluding the magnet and its case.
Figure\,\ref{fig:opt} shows these results for different magnet widths $W$ and heights $H$, while fixing the YBCO ring's cross-section of 28\,mm wide and 10\,mm high and setting \(h_{\mathrm{c}} = 6\,\mathrm{mm}\).
In general, larger magnets yield stronger forces, but also add weight to the rotor, producing a non-monotonic dependence of displacement on $W$ and $H$.
When performing simulations with different values of \(h_{\mathrm{c}}\), the absolute displacement changes, but the relative behavior with respect to the magnet dimensions remains largely unchanged.
Taking spatial allowance constraints into account as well, we adopted the dimensions of 22\,mm $\times$ 29\,mm, indicated by the star mark in Fig.\,\ref{fig:opt}, as the optimal design.

Furthermore, we conducted simulations with the introduction of a yoke, but no improvement was observed when considering the increase in rotor weight.
Although magnet arrangements such as the Halbach array can enhance the magnetic field in specific regions, we did not perform simulations for such configurations.
This is because our magnet ring is fabricated by first arranging the materials into a ring shape in an unmagnetized state and then magnetizing them using an external magnetic field, making it impossible to produce a ring with such an arrangement.

As for the YBCO, we found from simulation that lateral displacement is improved by about 28\% by transitioning from the previous design of circularly arranging 61 YBCO disks, each with a diameter of 28\,mm and height of 10\,mm,\cite{Yamada2024-ro} to a design featuring a full ring with a width of 28\,mm and height of 10\,mm.
Here, the force evaluation for the disk-shaped YBCO was conducted by dividing the disk into strip-shaped segments, simulating each separately, and summing their contributions.
For the ring-shaped YBCO and the magnet dimensions selected above, increasing the width or height of the YBCO beyond 28\,mm $\times$ 10\,mm did not result in significant improvements.

Based on these results, we adopted the combination of a 22\,mm $\times$ 29\,mm magnet and a 28\,mm $\times$ 10\,mm YBCO as the optimized design.
In this configuration, simulations predict that  if $h_{\mathrm{c}} = 6\,(5, 7)\,\mathrm{mm}$, the lateral stiffness is $2.0\times10^5\,(2.4\times10^5, 1.6\times10^5)\,\mathrm{N/m}$, resulting in a lateral displacement of 1.4 (1.1, 1.8)\,mm at $\theta = 50^\circ$.
This satisfies the requirement of less than 4\,mm with a safety factor of 2.8 (3.6, 2.2).

\subsection{Magnet Geometry Impact on Rotation}
\label{subsec:impact_rotation}

\begin{figure}
    \centering
    \includegraphics[width = 0.48\textwidth]{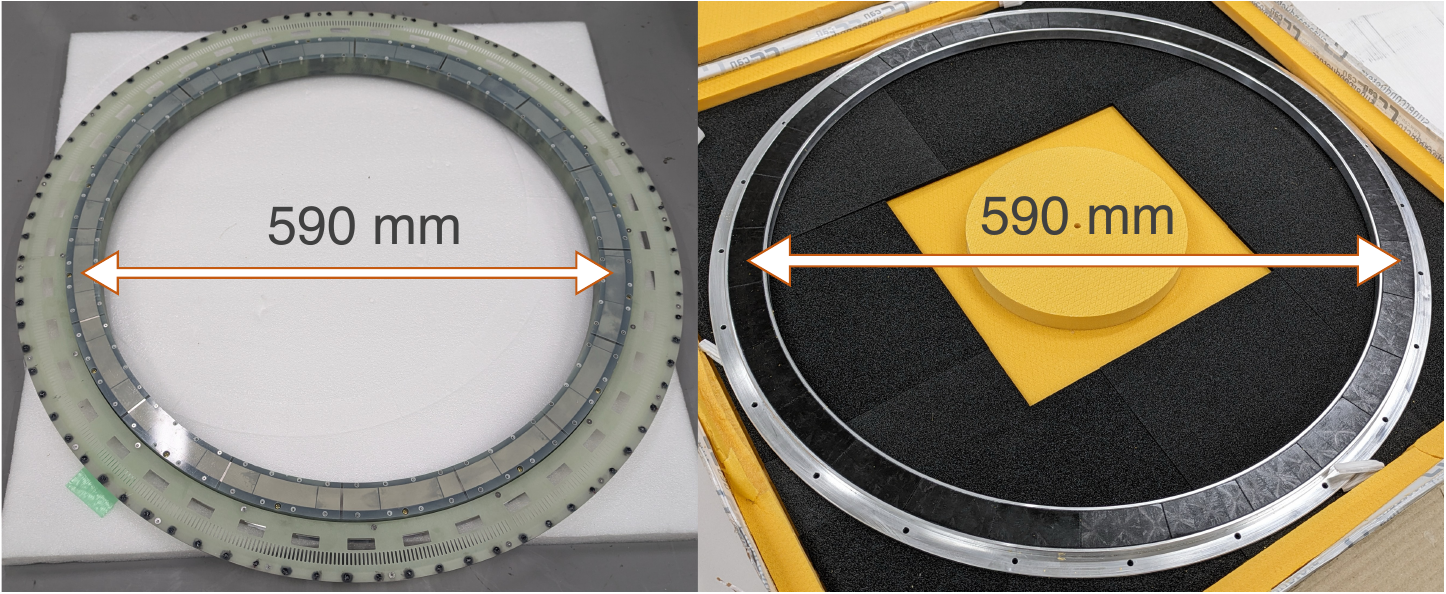}
    \caption{Left panel: Magnet assembly and encoder ring (rear view). Both the magnet case and the encoder ring are made of glass-epoxy. Right panel: YBCO assembly. A ring-shaped YBCO consisting of 31 segments is housed in an aluminum structure.}
    \label{fig:mag_ybco}
\end{figure}

Magnetic field uniformity is a critical factor for redusing frictional heat dissipation and enabling rotation at a small startup current.
The mechanisms of heat dissipation include hysteresis loss and eddy current loss, which depend on the cube and square of the magnetic field variation amplitude in the YBCO and conductor, respectively.\cite{Bean_1962, Bean_1964}
The larger magnetic field variation also results in a higher startup current due to the increased potential barrier in the rotational direction caused by flux pinning.
Here, we show that increasing the magnet size for stiffness is also favorable from the perspective of magnetic field uniformity.

From previous SMB development\cite{sakurai_estimation_2017, Sakurai2020, Yamada2024-ro}, we have found that (1) field non-uniformity due to variations among individual magnet segments is about 1\% of the magnetic field magnitude, and (2) small gaps ($\delta \sim 0.3$\,mm at room temperature and $\delta \sim 0.2$\,mm at cryogenic temperatures) between magnet segments produce larger field variations.\footnote{The glass-epoxy case and magnet materials have different thermal expansion coefficients, causing gaps to shrink upon cooling. A clearance of $\delta \sim 0.3$\,mm ensures that the magnets do not break at cryogenic temperatures.}

As shown in Appendix\,\ref{sec:delB_verification}, the segment gaps dominate the non-uniformity relevant to friction and startup.
We derive an approximate expression for the peak-to-peak field variation caused by a gap $\delta$ measured at a distance $h$ above the magnet surface (the derivation is provided in Appendix\,\ref{sec:deltaB}):
\begin{align}
\Delta B (W, H, \delta, h) = & \frac{\mu_0MW\delta}{2\pi} \Biggl(-\frac{1}{h\sqrt{4h^2 + W^2}} \notag \\
& + \frac{1}{\left(h + H\right)\sqrt{4\left(h + H\right)^2 + W^2}}\Biggr). \label{eq:deltaB}
\end{align}
Here, $M$ represents the magnetization of the magnet, and $\mu_0$ is the vacuum permeability.

To illustrate the effect of magnet size, consider two cases with $(W, H, h_{\mathrm{c}})$: Case\,1: $(15\,\mathrm{mm}, 15\,\mathrm{mm}, 5\,\mathrm{mm})$ and Case\,2: $(22\,\mathrm{mm}, 29\,\mathrm{mm}, 6\,\mathrm{mm})$, approximating previous SAT and current LF designs, respectively.
By force equilibrium with a load of $m = 35\,\mathrm{kg}$, $h_{\mathrm{l}}$ becomes 3.7,mm in Case\,1 and 4.9\,mm in Case\,2, with the corresponding lateral stiffness $K_{\mathrm{lat}}$ being $1.5\times10^5\,\mathrm{N/m}$ and $2.0\times10^5\,\mathrm{N/m}$, respectively.
In the following, we evaluate the magnetic field-related quantities at the YBCO position, $h = h_\mathrm{l}$, for the two cases.
Additionally, since $\delta \sim 0.3$,mm is determined based on machining precision and as a margin to prevent magnet breakage, it is independent of the size of the cross section.
Thus, we use the same value of $\delta$ for evaluating $\Delta B$ in both cases.
As a result, the magnetic field magnitude in Case\,2 increases to 115\% of that in Case\,1, whereas $\Delta B$ is reduced to 80\% of that in Case\,1.
In general, optimizing the magnet dimensions allows us to position the YBCO farther from the magnet while maintaining or enhancing stiffness, which in turn improves magnetic field uniformity (see Appendix\,\ref{sec:deltaB} for details).

\section{Implementation and Characterization}
\label{sec:impl_char}

\subsection{Implementation}
\label{subsec:impl}

\begin{figure}
    \centering
    \includegraphics[width = 0.35\textwidth]{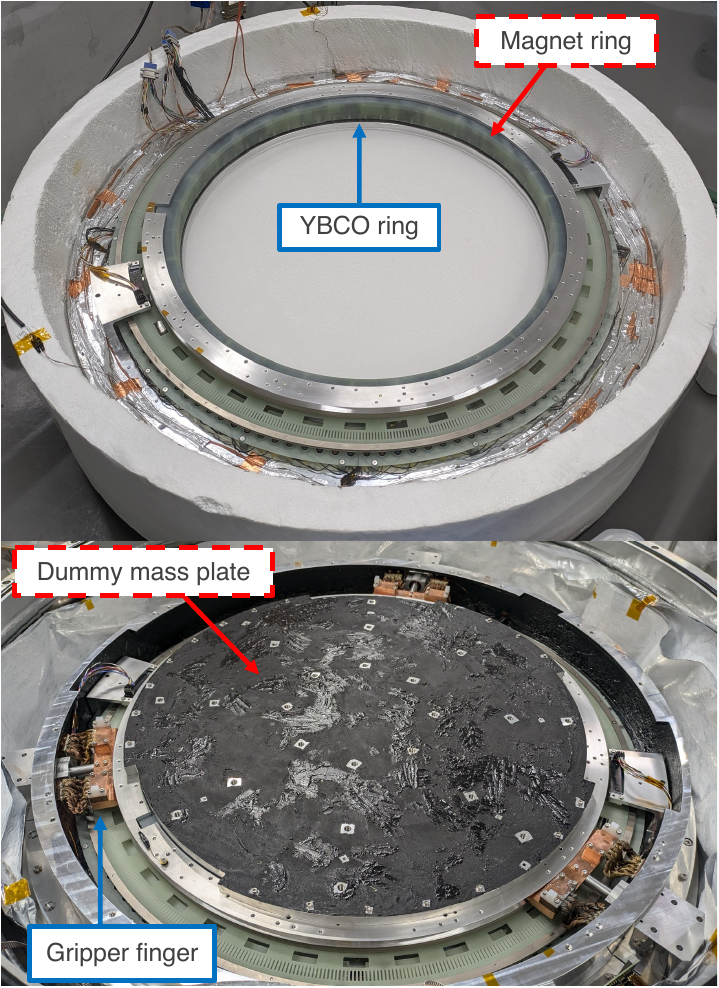}
    \caption{Top panel: The integrated SMB system during the liquid-nitrogen testing, placed inside a foam bucket containing liquid nitrogen. Dummy load plates can be placed on top of the rotor ring. Bottom panel: The SMB system installed in the cryostat, with aluminum dummy mass plates coated with epoxy added to replicate the load of the LF HWP.}
    \label{fig:assem}
\end{figure}

The left panel of Fig.\,\ref{fig:mag_ybco} shows the magnet assembly manufactured by Shin-Etsu Chemical\footnote{Shin-Etsu Chemical Co., Ltd., \protect\url{https://www.shinetsu.co.jp/en/}}, consisting of 31 segments with optimal geometry (i.e., indicated by a star in Fig.\,\ref{fig:opt}).
These segments are secured in a glass-epoxy case.
The right panel of Fig.\,\ref{fig:mag_ybco} shows the YBCO assembly fabricated by CAN Superconductors\footnote{Can Superconductors, \protect\url{https://www.can-superconductors.com/}}, consisting of 29 YBCO tile segments grown from three seeds each, arranged to form a ring with a 28\,mm $\times$ 10\,mm cross-section.
The YBCO segments are housed in an aluminum structure, and the entire assembly is mounted on a glass-epoxy base.
The number of magnet and YBCO segments are both prime numbers.
This approach is adopted because it is known that significant vibration modes can occur at frequencies that are the rotor's rotation frequency multiplied by common multiples of the segment numbers.\cite{Yamada2024-ro}
By using prime numbers for both segment counts, we significantly increase this common multiple, effectively eliminating the impact of such vibration modes.

In our design, the total weight of the magnets, their case, the encoder ring, and the aluminum part for mounting the load, denoted as \(m_{\mathrm{fixed}}\), is 13.2\,kg.
Hereafter, $m_{\mathrm{load}}$ refers to the weight of the load mounted on these structures, which corresponds to the weight of the sapphire stack optic and its supporting structure.
The rotor mass, $m$, represents the total weight of the rotor, given by $m = m_{\mathrm{fixed}} + m_{\mathrm{load}}$.

We integrated the magnet assembly, the YBCO assembly and other components into a complete SMB system and performed measurements of stiffness and rotation both in liquid-nitrogen and in a cryostat.

\subsection{Stiffness Measurement}
\label{subsec:stiff_meas}

\begin{figure}
    \centering
    \includegraphics[width = 0.4\textwidth]{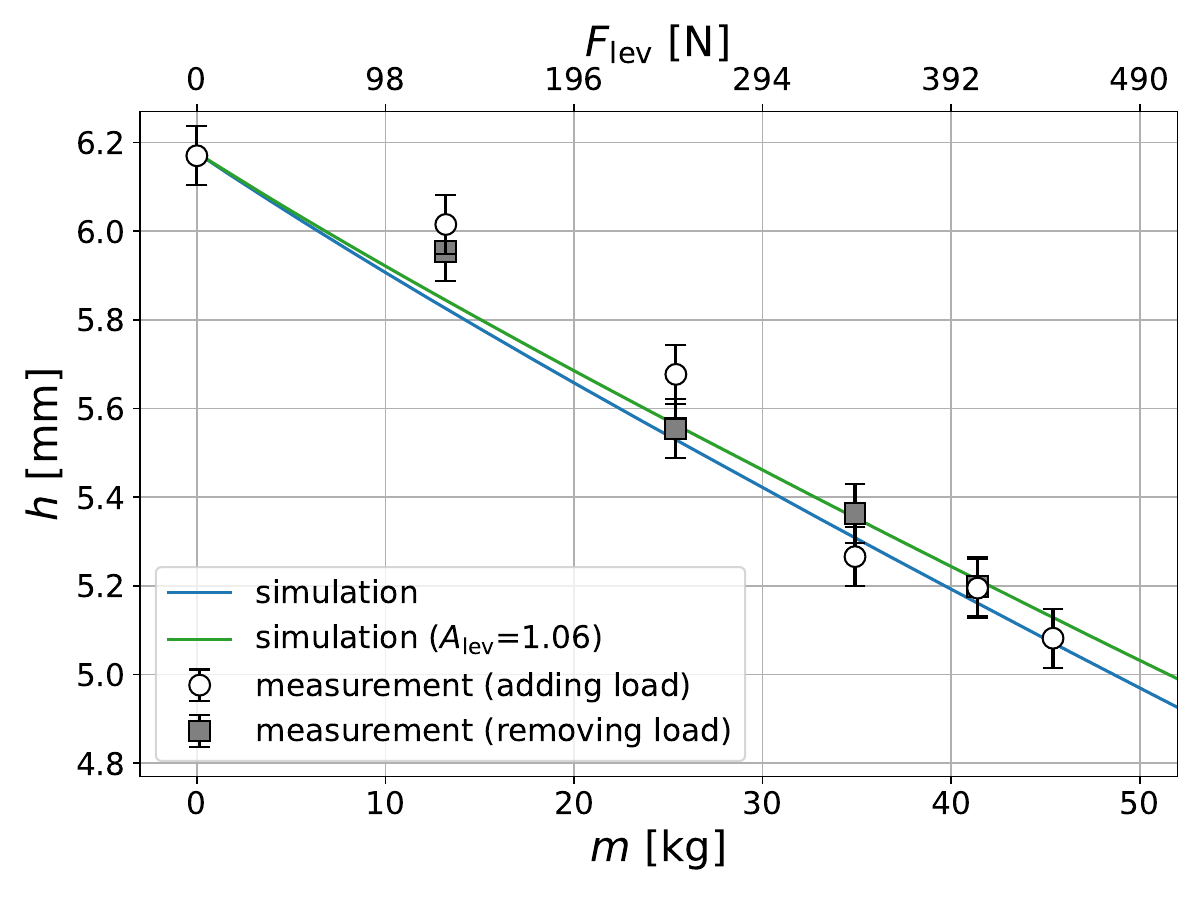}
    \caption{Measured rotor height \(h\) with various rotor mass \(m\). The point at \(m = 0\,\mathrm{kg}\) corresponds to the measurement with spacer jigs in place, giving \(h_\mathrm{c}=6.17\,\mathrm{mm}\). Open circles show data while incrementally adding loads, and the filled squares represent data taken while removing loads gradually from the maximum load. The blue curve is the simulation, while the green curve is the simulation scaled by 1.06.}
    \label{fig:vsag}
\end{figure}

We measured vertical displacement of the rotor for various \(m_\mathrm{load}\) during the tests in liquid-nitrogen (see the top panel of Fig.\,\ref{fig:assem}).
During field cooling in liquid-nitrogen, $h_{\mathrm{c}}\simeq 6\,\mathrm{mm}$ was set using spacer jigs, which were subsequently removed for levitation.
We then successively added and removed aluminum dummy load plates on the rotor.
We measured the rotor height $h$ while the jigs were in place and for each loaded condition using a coordinate measuring machine (CMM).
Figure\,\ref{fig:vsag} shows these measurements along with our simulation.
To compare data and simulation, we scale the simulated levitation force by a factor $A_{\mathrm{lev}}$: \(F_\mathrm{lev}' = A_{\mathrm{lev}}F_{\mathrm{lev}}\). 
A fit to the data yields $A_{\mathrm{lev}} = 1.06\pm0.03$, suggesting our vertical force prediction is accurate to within 6\%. 
At \(m_\mathrm{load} = 32\,\mathrm{kg}\), (close to the LF HWP mass), the vertical displacement is 1.1\,mm and the vertical stiffness is $4.4\times10^5\,\mathrm{N/m}$.

\begin{table}
\caption{Configurations in each cryostat run. Three runs were conducted with different rotor masses and cooling heights. The cooling height \(h_{\mathrm{c}}\) was measured with a CMM. The levitation height $h_{\mathrm{l}}$ presented here is the value at $\theta = 90^\circ$ and is estimated from the measured $h_{\mathrm{c}}$ and the vertical force simulation.}
\centering
\begin{tabular}{c|ccc}
        & $m\ (m_\mathrm{load})$ {[}kg{]} & $h_{\mathrm{c}}$ {[}mm{]} & $h_{\mathrm{l}}$ {[}mm{]} \\ \hline
Run 1 & $25.4\ (12.2)$                & $5.07\pm 0.07$                    & $4.55\pm 0.07$                  \\
Run 2 & $45.2\ (32.0)$                & $4.93\pm 0.07$                   & $4.05\pm 0.07$                  \\
Run 3 & $45.2\ (32.0)$                & $5.88\pm 0.07$                    & $4.81\pm 0.07$                 
\end{tabular}
\label{table:configuration}
\end{table}

\begin{figure}
    \centering
    \includegraphics[width = 0.35\textwidth]{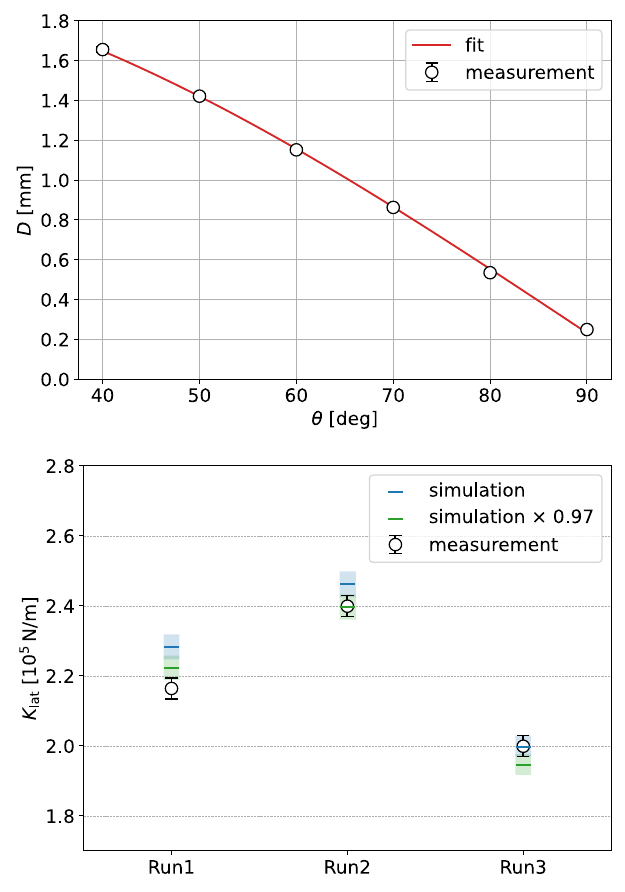}
    \caption{Top panel: Lateral displacement $D$ versus elevation angle $\theta$ during Run\,2. The red line is a fit using Eq.\,\eqref{eq:displacement} plus a small offset at $\theta = 90^\circ$. Bottom panel: Measured lateral stiffness for Runs\,1-3. Blue lines are simulations, and green lines are simulations scaled by $A_{\mathrm{lat}} = 0.97$. The shaded region reflect simulation uncertainty inherited from $h_{\mathrm{c}}$ measurement uncertainty.}
    \label{fig:lateral}
\end{figure}

We measured lateral displacement and lateral stiffness in three cryostat runs (Table\,\ref{table:configuration}) at different rotor masses and cooling heights.
We varied the rotor mass by changing the dummy mass plates and altered the cooling height by modifying height of the gripper fingers.
In each run, after field cooling at $\theta = 90^\circ$, cryostat was tilted to elevations between 40 to 90\,degrees.
The rotor displacement was measured by the pair of optical encoders, floowing the method described in \citet{Yamada2024-ro}.

The top panel of Fig.\,\ref{fig:lateral} shows $D$ vs. $\theta$ from Run\,2 as an example..
We fit the data using Eq.\,\eqref{eq:displacement} with an added offset at $\theta = 90^\circ$ to account for the initial misalignment.
Although $K_{\mathrm{lat}}$ formally depends on $\theta$, this variation is small (Appendix\,\ref{sec:klat}).
Hence, we treat $K_{\mathrm{lat}}$ as a single best-fit constant in each run, and the obtained value is interpreted as the one at the intermediate elevation, specifically at $\theta = 60^\circ$, for comparison to simulation.

The bottom panel of Fig.\,\ref{fig:lateral} summarizes $K_{\mathrm{lat}}$ measurements from all runs.
For comparison, the simulation for $K_{\mathrm{lat}}$ at $\theta = 60^\circ$ is also shown.
By fitting the measurements to simulations uniformly scaled by a factor $A_{\mathrm{lat}}$, we obtain $A_{\mathrm{lat}} = 0.97\pm0.01$.
This indicates 3\% agreement between simulation and data.
In Run\,3, which closely matches actual operating conditions, $K_{\mathrm{lat}} = 2.0\times10^5\,\mathrm{N/m}$.
The lateral displacement is 1.4\,mm at $\theta = 50^\circ$, well below the 4\,mm requirement.

In summary, we measured the vertical and lateral stiffness of the SMB to be $4.4\times10^5\,\mathrm{N/m}$ and $2.0\times10^5\,\mathrm{N/m}$, respectively, with $h_{\mathrm{c}}\sim 6\,\mathrm{mm}$ and a load equivalent to SAT-LF HWP.
The SMBs in SAT-MFs have a vertical stiffness of approximately $1.7\times10^5\,\mathrm{N/m}$ and a lateral stiffness of about $0.8\times10^5\,\mathrm{N/m}$.\cite{Yamada2024-ro}
This demonstrates that our optimized design has achieved an improvement of factor 2.5.

\subsection{Rotation Tests}
\label{subsec:rotation}

\begin{figure}
    \centering
    \includegraphics[width = 0.40\textwidth]{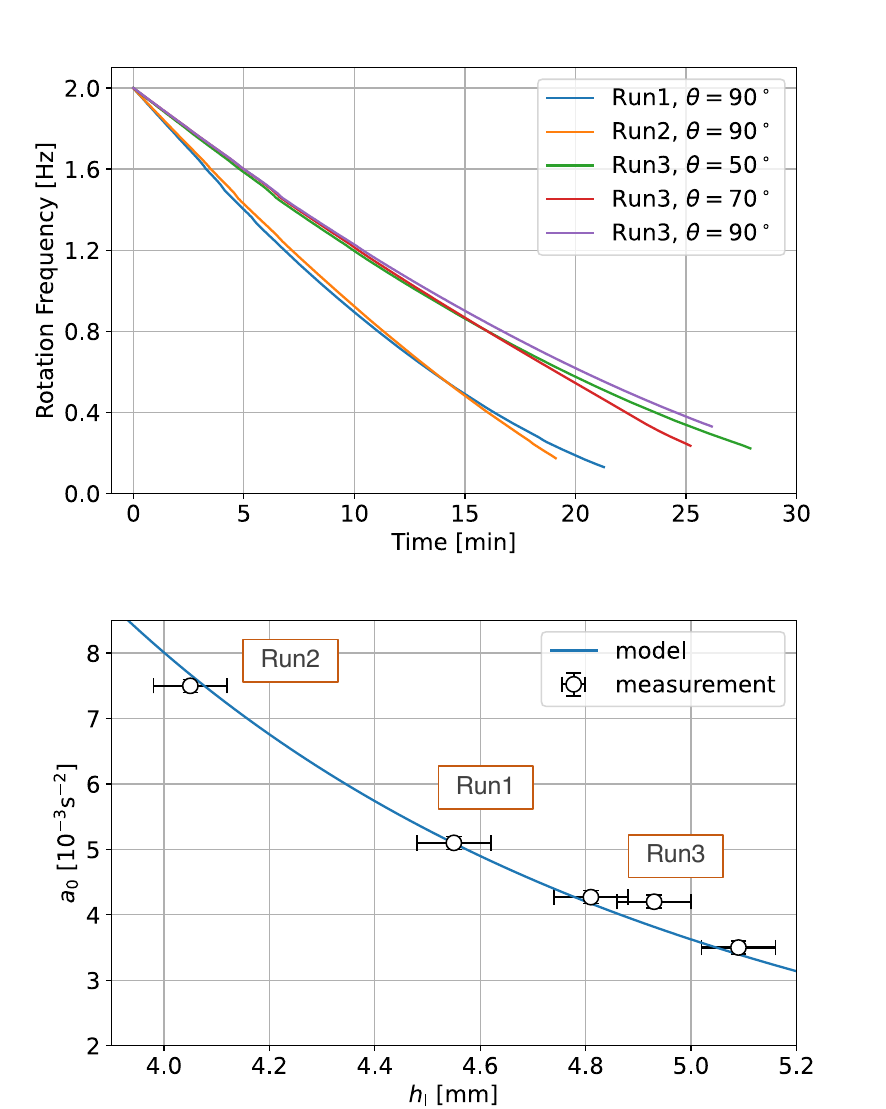}
    \caption{Top panel: Rotor frequency over time in spin-down measurements. Data were taken at $\theta = 90^\circ$ for Run\,1 and Run\,2, while Run\,3 measurements were done at three different elevations. Bottom panel: Hysteresis loss coefficients $a_0$ for each dataset, derived from fitting Eq.\,\eqref{eq:spin-down}. The horizontal axis is the levitation height \(h_\mathrm{l}\), determined by \(m\), \(h_{\mathrm{c}}\), \(\theta\). The circles, from left to right, correspond to \(\theta = 90^\circ\) in Run\,1, \(\theta = 90^\circ\) in Run\,2, and \(\theta = 90^\circ, 70^\circ,\) and \(50^\circ\) in Run\,3. The blue curve is a fit proportional to \((\Delta B)^3\) calculated from Eq.\,\eqref{eq:deltaB}.}
    \label{fig:friction}
\end{figure}

We conducted rotation tests in the same three cryostat runs (Table\,\ref{table:configuration}), successfully achieving continuous rotation at 2 Hz in all cases.
Runs\,2 and 3 had $m_{\mathrm{load}} = 32$\,kg, similar to the SAT-LF HWP, but different cooling heights.
Below, we discuss the friction and startup performance.

In each run, we performed spin-down measurements by stopping the rotation drive near 2\,Hz and letting the rotation freely decay.
The top panel of Fig.\,\ref{fig:friction} shows examples of the rotor frequency $f$ as a function of time.
We fit the frequency decay using:\cite{Hull_2000, Sakurai_2017}
\begin{equation}\label{eq:spin-down}
-2\pi\frac{df}{dt} = a_0+2\pi a_1 f,
\end{equation}
where $a_0$ is the hysteresis loss coefficient and $a_1$ is the eddy current loss coefficient.

The bottom panel of Fig.\,\ref{fig:friction} shows $a_0$ vs. the levitation height \(h_{\mathrm{l}}\) for each dataset.
As expected, a higher levitation improves uniformity of magnetic field on the YBCO, reducing the hysteresis loss.
Comparing the measurements from Run\,2 and Run\,3 at $\theta = 90^\circ$, we find that a difference of approximately 0.8\,mm in \(h_{\mathrm{l}}\) resulted in a twofold improvement in $a_0$.

According to Bean's model, hysteresis loss is proportional to $(\Delta B)^3$.\cite{Bean_1962, Bean_1964}
Evaluating Eq.\,\eqref{eq:deltaB} with $\delta = 0.2\,\mathrm{mm}$, $W = 22\,\mathrm{mm}$, and $H = 29\,\mathrm{mm}$, we find the data are well fit by a simple scaling of $(\Delta B)^3$, as shown by blue curve in the bottom panel of Fig.\,\ref{fig:friction}.
This agreement supports the validity of our magnetic field non-uniformity model.

For eddy current loss, we observed no significant dependence on \(h_{\mathrm{l}}\), with $a_1$ estimated to be $3.8\times10^{-4}\,\mathrm{s}^{-1}$.
This lack of variation is reasonable for two reasons: first, eddy current loss depends on $(\Delta B)^2$ rather than $(\Delta B)^3$, and second, the metal components causing the eddy current loss are farther from the magnet compared to the YBCO, resulting in a small fractional difference in the distance between them with changes in the levitation height.

Heat dissipation of the SMB due to friction can be estimated using the formula $I(a_0+2\pi a_1f)2\pi f$ where $I$ is the moment of inertia of the rotor.
In Run,3, with the moment of inertia evaluated as $I = 2.5\,\mathrm{kg}\cdot\mathrm{m}^2$, the heat dissipation during 2\,Hz rotation at $\theta=90^\circ$ is estimated to be approximately 0.26\,W.
This consists of 0.11,W from hysteresis loss and 0.15,W from eddy current loss.
In SAT-MF, the heat dissipation is reported to be about 0.15\,W \cite{Yamada2024-ro}, showing no significant increase compared to the total heat dissipation requirement of 3\,W for the polarization modulator.
These results confirm that the magnetic field uniformity is sufficient to satisfy the heat dissipation requirement.

The improved magnetic field uniformity at larger $h_{\mathrm{l}}$ also makes the initiation of rotation easier with reduced potential energy barrier.
In Run\,3, rotation was successfully initiated with total current under 1\,A, satisfying the requirement, ahereas Run\,2 required $>1\,\mathrm{A}$.

Thus, we choose $h_{\mathrm{c}}\sim 6\,\mathrm{mm}$ for normal SAT-LF operation to benefit both alignment and rotation requirements.

\section{Conclusion}
\label{sec:con}

We have designed and characterised an SMB for the polarization modulator in SO's SAT-LF, featuring a clear aperture larger than 550\,mm and a load capacity exceeding 35\,kg.
The key innovation of this study is a new design methodology based on finite element simulations, which guided us in optimizing the geometry of the magnet and the superconductor.
The measurements of force and stiffness demonstrated that our SMB meets the alignment requirements and that our model successfully predicts these quantities at the 5\% accuracy level.
Furthermore, rotation tests confirmed that the friction is sufficiently low and that the required current to initiate rotation is small, verifying that the rotational requirements are satisfied.

We will now proceed with the development of the 30/40\,GHz band HWP, integrating it with the SMB developed in this study to build the complete SAT-LF polarization modulator.
This will be the first realization of a cryogenic continuously rotating HWP-based polarization modulator in this frequency band.
Observations of the SAT-LF with this new modulator will further advance the search for primordial B-modes in the CMB.

Future experiments, such as CMB-S4\cite{abazajian2019cmbs4sciencecasereference}, will likely require even larger and stronger SMBs.
The design methodology and implementation of the SMB demonstrated here represent a crucial stepping stone for such future projects.

\begin{acknowledgments}
This work was supported by JSPS KAKENHI Grant Numbers 22H04913, 23H00105, and 23K25898, and by the JSPS Core-to-Core program Grant Number JPJSCCA20200003.
This work was supported in part by a grant from the Simons Foundation (Award \#457687, B.K.) and the National Science Foundation (Award Number: 2153201).
D.S. acknowledges the support from FoPM, WINGS Program, the University of Tokyo. 
J.S. acknowledges the support from the International Graduate Program for Excellence in Earth-Space Science (IGPEES) and the JSR Fellowship, the University of Tokyo. 
We thank Hiroko Tomoda for providing valuable insights by conducting a small-scale experiment prior to this development.
We thank John Groh and Johannes Hubmayr for their valuable comments and suggestions on the manuscript.
\end{acknowledgments}

\tocless{\section*{Author Declarations}}

\tocless{}{\subsection*{Conflict of Interest}}
The authors have no conflicts to disclose.\\
\tocless{}{\subsection*{Author Contributions}}
\noindent \textbf{D.\,Sasaki}: Conceptualization (equal); Investigation (lead); Methodology (equal); Writing - original draft (lead).
\textbf{J.\,Sugiyama}: Investigation (supporting); Writing - original draft (supporting).
\textbf{K.\,Yamada}: Investigation (supporting); Writing - review \& editing (supporting).
\textbf{B.\,Bixler}: Investigation (supporting); Writing - review \& editing (supporting).
\textbf{Y.\,Sakurai}: Conceptualization (equal); Funding acquisition (supporting); Investigation (supporting); Methodology (equal); Project administration (equal).
\textbf{K.\,Arnold}: Funding acquisition (supporting); Project administration (equal); Supervision (supporting).
\textbf{B.\,R.\,Johnson}: Writing - review \& editing (equal).
\textbf{A.\,Kusaka}: Funding acquisition (lead); Project administration (equal);  Supervision (equal); Writing - original draft (supporting).
\\
\tocless{\section*{Data availability}}
The data that support the findings of this study are available from the corresponding author upon reasonable request.

\appendix
\section{Dependence of \(K_\mathrm{lat}\) on the Elevation Angle}
\label{sec:klat}

\begin{figure}[h]
    \centering
    \includegraphics[width = 0.48\textwidth]{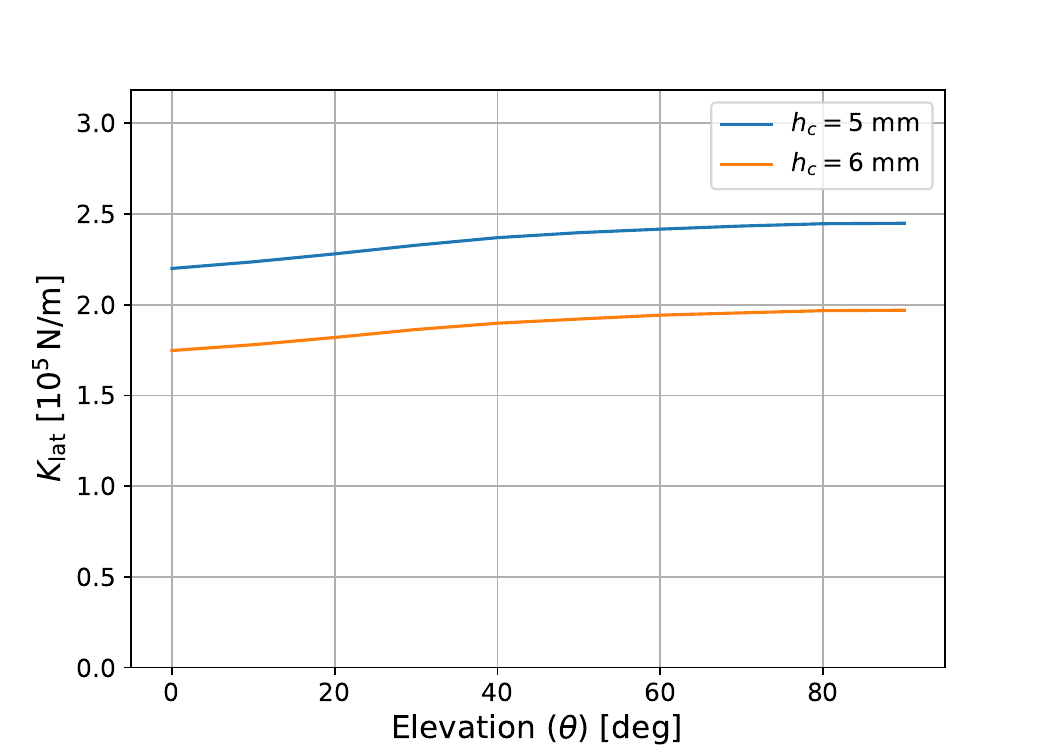}
    \caption{Simulation of $K_{\mathrm{lat}}$ vs. $\theta$ for the optimized magnet-YBCO design. The blue curve is the results for $h_{\mathrm{c}}=6\,\mathrm{mm}$, and the orange curve is the results for $h_{\mathrm{c}}=5\,\mathrm{mm}$. From $40^\circ$ to $90^\circ$, the variation is only a few percent.}
    \label{fig:klat_theta}
\end{figure}

As described in Sec.\,\ref{subsec:sim}, $K_{\mathrm{lat}}$ depends on the telescope elevation angle $\theta$ via $h_{\mathrm{l}}$ from $F_{\mathrm{lev}}(h_{\mathrm{l}}) = mg\sin\theta$. 
Figure\,\ref{fig:klat_theta} shows the simulated $K_{\mathrm{lat}}$ vs.\ $\theta$ for our optimized design.
In the range of interest ($40^\circ<\theta<90^\circ$), $K_{\mathrm{lat}}$ remains almost constant with a variation of approximately 2\%.
At $\theta=0^\circ$, $K_{\mathrm{lat}}$ is still $\sim90\%$ of its value at $90^\circ$.

\section{Derivation of Magnetic Field Non-Uniformity Due to Gaps}
\label{sec:deltaB}

\begin{figure}[h]
    \centering
    \includegraphics[width = 0.35\textwidth]{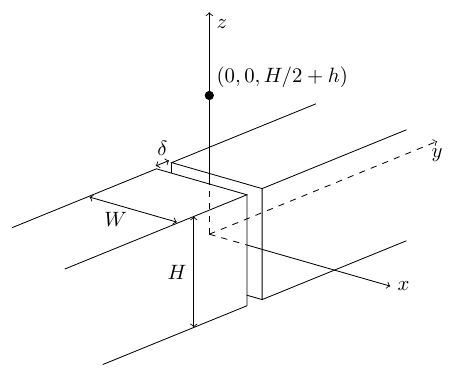}
    \caption{Diagram of the two semi-infinite magnets with a gap $\delta$ between them. }
    \label{fig:tikz}
\end{figure}

Here, we derive the magnetic field variation $\Delta B$ at a distance $h$ from the surface of the magnet, caused by the gap $\delta$ between the magnets.
As shown in Fig.\,\ref{fig:tikz}, we consider two semi-infinite magnets with a width $W$ and a height $H$.
The two magnets face each other in the $zx$-plane, separated by a gap of $\delta$.
The centers of the facing surfaces are located at $(0, \delta/2, 0)$ and $(0, -\delta/2, 0)$, respectively.
The magnets are magnetized in the $z$-direction, with a magnetization magnitude of $M$.
We define and calculate $\Delta B$ as the difference in the $z$-direction magnetic field at the point $(0, 0, H/2 + h)$, between the cases where a gap $\delta$ is present and absent.
This $\Delta B$ is a good approximation of the the peak-to-peak amplitude of the magnetic field variation along the $y$-axis measured at $h$ above the magnet surface when $\delta$ is significantly smaller than the length of the magnet segment, which is true in our case.

First, we consider the expression for the magnetic field generated by a rectangular magnet with uniform magnetization.
The magnetic scalar potential $\phi_m$ due to magnetization $M$ is expressed in terms of the magnetic pole density $\rho_m$ as follows:
\begin{equation}\label{eq:phi_m}
\phi_m(\bm{r}) = \frac{1}{4\pi} \int \frac{\rho_m(\bm{r'})}{|\bm{r} - \bm{r'}|} \, dV'.
\end{equation}
For uniform magnetization, the magnetic pole density inside the volume is zero, while a surface magnetic pole density $\sigma_m$ exists.
The density $\sigma_m$ is given by the dot product of the surface normal vector $\hat{n}$ and the magnetization $\bm{M}$:
\begin{equation}\label{eq:sigma_m}
\sigma_m = \bm{M} \cdot \hat{\bm{n}}.
\end{equation}
In the configuration considered here, where the magnetization is in the $z$ direction, the surface magnetic pole density $\sigma_m$ appears only on the surfaces at $z = \pm H/2$ and is given by $\sigma_m = M \equiv |\bm{M}|$.
Thus, in the absence of a gap, the magnetic scalar potential $\phi_m$ is given by:
\begin{align}
    &\phi_m (x, y, z) \notag \\ 
    &=  \frac{M}{4\pi} \int_{-W/2}^{W/2}dx'\int_{-\infty}^{\infty}dy' \Biggl[\frac{1}{\sqrt{(x - x')^2 + (y - y')^2 + (z - H/2)^2}}  \notag \\
    & \qquad \qquad -  \frac{1}{\sqrt{(x - x')^2 + (y - y')^2 + (z + H/2)^2}}\Biggr]. \label{eq:phi_m_no_gap}
\end{align}
The $z$-component of the magnetic field is obtained by taking the negative derivative of $\phi_m$ with respect to $z$:
\begin{align}
    &B_z (0, 0, H/2+h)\notag \\ 
    &= \frac{\mu_0M}{4\pi} \int_{-W/2}^{W/2} dx'\int_{-\infty}^{\infty}dy'\Biggl[ \frac{h}{\left[x'^2 + y'^2 + h^2\right]^{3/2}} \, dx' \, dy' \notag \\
    & \qquad \qquad \qquad - \frac{(H + h)}{\left[x'^2 + y'^2 + (H + h)^2\right]^{3/2}} \Biggr]. \label{eq:B_no_gap}
\end{align}
By integrating over $x'$ and $y'$, we obtain the following expression:
\begin{align}
B_z(&0, 0, H/2+h) \notag \\ & = \frac{\mu_0M_z}{\pi}\left(-\arctan\left(\frac{W}{2h}\right) 
+\arctan\left(\frac{W}{2(h+H)}\right)\right) \label{eq:B} \\
& \equiv B(W, H, h). \notag
\end{align}

When a gap is present, the integration range for $y'$ along $(-\infty, \infty)$ is split into two intervals: $(-\infty, -\delta/2)$ and $(\delta/2, \infty)$.
Thus, the difference in the magnetic field between the cases with and without a gap is given by:
\begin{align}
    &\Delta B (W, H, \delta, h) \notag \\
    &=  \frac{\mu_0M}{4\pi} \int_{-W/2}^{W/2} dx'\int_{-\delta/2}^{\delta/2}dy'\Biggl[ \frac{-h}{\left[x'^2 + y'^2 + h^2\right]^{3/2}} \notag \\
    & \qquad \qquad \qquad + \frac{(H + h)}{\left[x'^2 + y'^2 + (H + h)^2\right]^{3/2}} \Biggr] \notag \\
    &\simeq \frac{\mu_0M\delta}{4\pi} \int_{-W/2}^{W/2} dx'\Biggl[ \frac{-h}{\left[x'^2 + h^2\right]^{3/2}} + \frac{(H + h)}{\left[x'^2 + (H + h)^2\right]^{3/2}} \Biggr]. \label{eq:deltaB_int}
\end{align}
Here, since $\delta \ll W, H, h$, an approximation was made by neglecting the $y'$ term in the denominator when obtaining the second line. 
By performing the integration over $x'$, we can obtain Eq.\,\eqref{eq:deltaB},
\begin{align}
\Delta B (W, H, \delta, h) = & \frac{\mu_0MW\delta}{2\pi} \Biggl(-\frac{1}{h\sqrt{4h^2 + W^2}} \notag \\
& + \frac{1}{\left(h + H\right)\sqrt{4\left(h + H\right)^2 + W^2}}\Biggr). \label{eq:deltaB_app}
\end{align}

\begin{figure}[h]
    \centering
    \includegraphics[width = 0.35\textwidth]{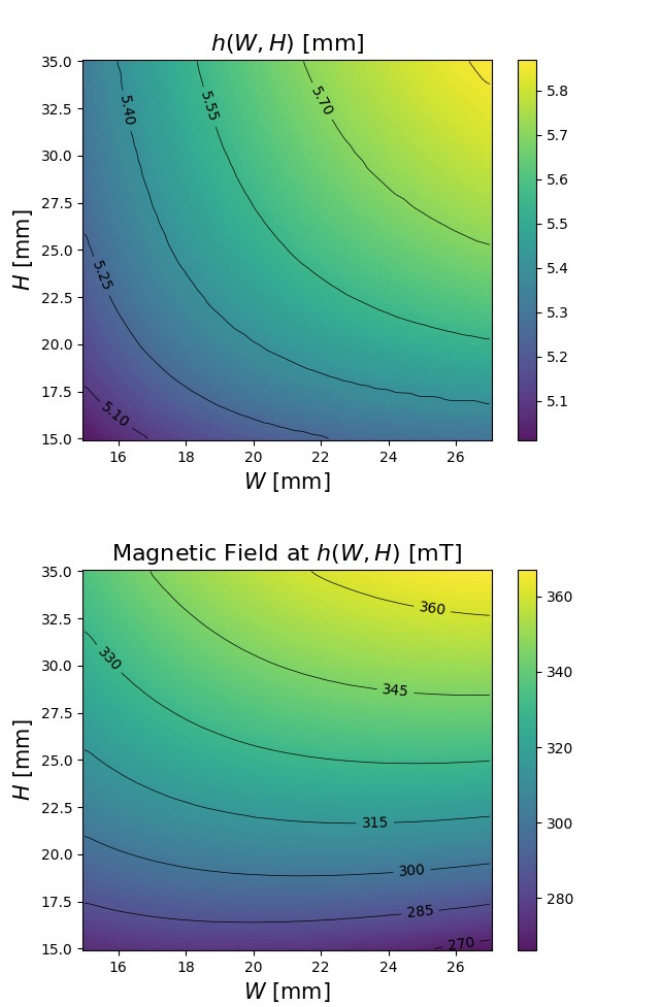}
    \caption{Top panel: A color plot of $h(W, H)$ as a function of magnet width ($W$) and height ($H$), where $\Delta B(W, H, \delta, h(W, H)) = \Delta B(W = 15\,\mathrm{mm}, H = 15\,\mathrm{mm}, \delta, h = 5\,\mathrm{mm}) = \mathrm{const}$. When the magnet size increases, the field fluctuations remain constant if the measurement position is moved farther from the magnet. Bottom panel: The magnetic field magnitude evaluated using $B(W, H, h(W, H))$, based on the $h(W, H)$ values obtained in the top panel. While $\Delta B$ remains constant, the magnetic field increases with the size of the magnet, as shown in the plot.}
    \label{fig:B_app}
\end{figure}

Eq.\,\eqref{eq:B} and Eq.\,\eqref{eq:deltaB_app} exhibit different dependencies on $W$, $H$, and $h$.
Therefore, as shown in Fig.\,\ref{fig:B_app}, by appropriately adjusting $h$ in response to changes in $W$ and $H$, it is possible to increase the magnetic field while keeping $\Delta B$ constant.
This suggests that as long as the magnetic field variation caused by the gaps remains dominant, increasing the size of the magnets allows for a stronger magnetic field at the YBCO position without increasing the field variation.
There is a general trend that the larger magnet size and the stronger magnetic field on YBCO lead to the larger $K_{\mathrm{lat}}$, which in turn reduces the displacement.
For rigorous verification of the stiffness increase, finite element simulation should be performed for each specific configuration.

\section{Verification of the Magnetic Field Non-Uniformity Model}
\label{sec:delB_verification}

\begin{figure}[h]
  \centering
  \includegraphics[width=0.8\linewidth]{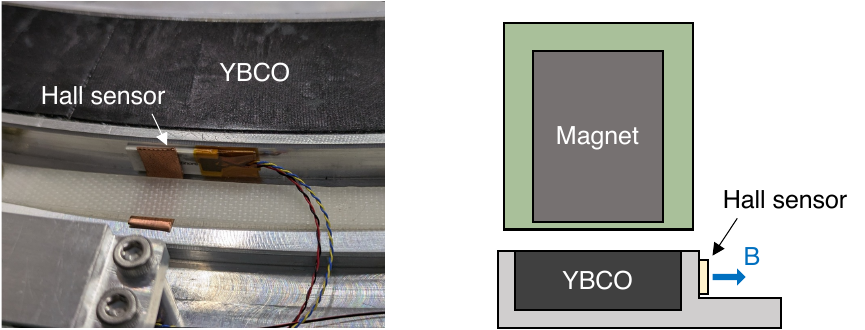}
  \caption{Left panel: Placement of the Hall sensors for measuring the magnetic field during the rotor's liquid-nitrogen test. Right panel: Schematic of the hall sensor's measurement position. Blue arrow indicates the sensitivity direction of the magnetic field.}
  \label{fig:mag_meas}
\end{figure}

In this section, we verify the model of magnetic field non-uniformity caused by segment gaps by comparing it to our measurements. 
During the liquid-nitrogen test described in Sec.\,\ref{subsec:stiff_meas}, Hall sensors were placed near the YBCO as shown in Fig.\,\ref{fig:mag_meas}. 
While the rotor was in motion, the magnetic field was measured at a sampling rate of  4000\,Hz and recorded along with the rotor angle from the optical encoder, allowing us to reconstruct the magnetic-field strength as a function of the rotor angle.
Note that the rotor magnet is not immersed in the liquid nitrogen during this liquid nitrogen test.
Thus, the rotor magnet was at an intermediate temperature between room temperature and liquid-nitrogen temperature.

\begin{figure}[t]
  \centering
  \includegraphics[width=0.9\linewidth]{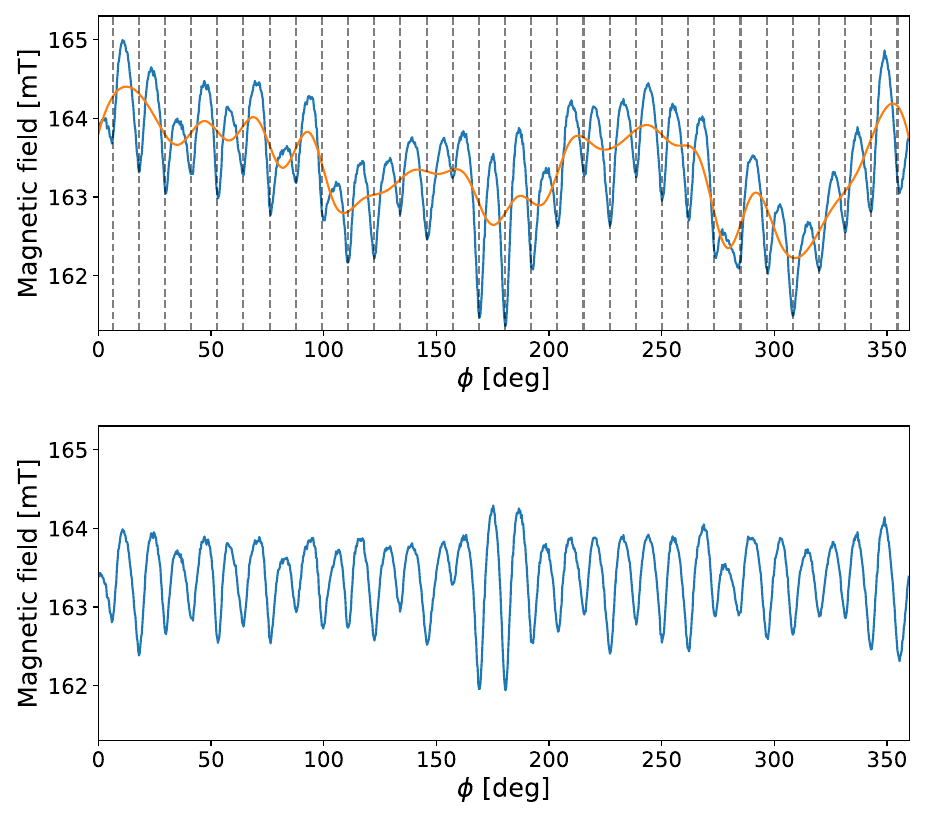}
  \caption{Upper panel: Measured magnetic-field strength as a function of rotor angle (blue), overlaid with a smooth component due to individual magnet variations (orange). The orange curve was obtained by smoothly connecting the 31 points, each representing the average magnetic field calculated for each segment as indicated by the gray dashed lines. Lower panel: The residual after subtracting the smooth component, showing periodic features arising from the 31 segment gaps.}
  \label{fig:mag_data}
\end{figure}

The blue curve in the upper panel of Fig.\,\ref{fig:mag_data} represents the measured data.
The superposed orange curve is obtained by calculating the average magnetic field for each magnet segment interval and smoothly interpolating the resulting 31 points.
This smooth variation originates from differences among individual segments, with a peak-to-peak amplitude of approximately 1\% of the total magnetic field strength.
By subtracting this smooth component from the measured data, we isolate the contribution from the segment gaps, as shown in the lower panel of Fig.\,\ref{fig:mag_data}.

Bean's model\cite{Bean_1962, Bean_1964} assumes that the hysteresis in the magnetization of superconductors is described by the square of the magnetic field, leading to energy loss proportional to $(\Delta B)^3$ per field loop.
Extending this idea to our system, the total loss per rotor revolution is proportional to the sum of the cubed peak-to-peak magnetic field variations $\Delta B_{\mathrm{pp},I}$ over each monotonically increasing or decreasing interval $I$:
\begin{equation}
a_0 \propto \sum_{\mathrm{monotonic\ intervals\ } I} |\Delta B_{\mathrm{pp},I}|^3.
\end{equation}

On the other hand, since eddy current loss is proportional to the square of the time derivative of the magnetic field in the conductor, we obtain: 
\begin{equation} 
a_1 \propto \int_0^{2\pi} \left( \frac{dB}{d\phi} \right)^2 d\phi. 
\end{equation}

By numerically evaluating these values using the magnetic field data from the upper and lower panels of Fig.\,\ref{fig:mag_data}, we estimate the contributions of segment-to-segment variations and segment gaps to the loss.
As a result, we find that $a_0$ and $a_1$ in the lower panel are 88\% and 96\% of those in the upper panel, respectively.
This indicates that magnetic field variations caused by segment gaps make a dominant contribution to the loss, rather than the smoother and longer-period variations due to individual segment differences.

Since $\Delta B(W, H, \delta, h)$ in Eq.\,\eqref{eq:deltaB_app} represents the peak-to-peak field variation caused by segment gaps, the following approximation holds:
\begin{align}
    a_0 &\propto N_{\mathrm{seg}}\left(\Delta B(W, H, \delta, h_{\mathrm{l}})\right)^3, \\
    a_1 &\propto N_{\mathrm{seg}}\left(\Delta B(W, H, \delta, h)\right)^2,
\end{align}
where $N_{\mathrm{seg}}$ is the number of segments.
This framework explains the successful model fitting shown in Fig.\,\ref{fig:friction}.
As indicated by these results, in addition to the strategy we adopted of increasing the levitation height while maintaining stiffness, minimizing $N_{\mathrm{seg}}$ is also an effective approach in addressing the issue of heat dissipation.

\begin{figure}[t]
  \centering
  \includegraphics[width=0.8\linewidth]{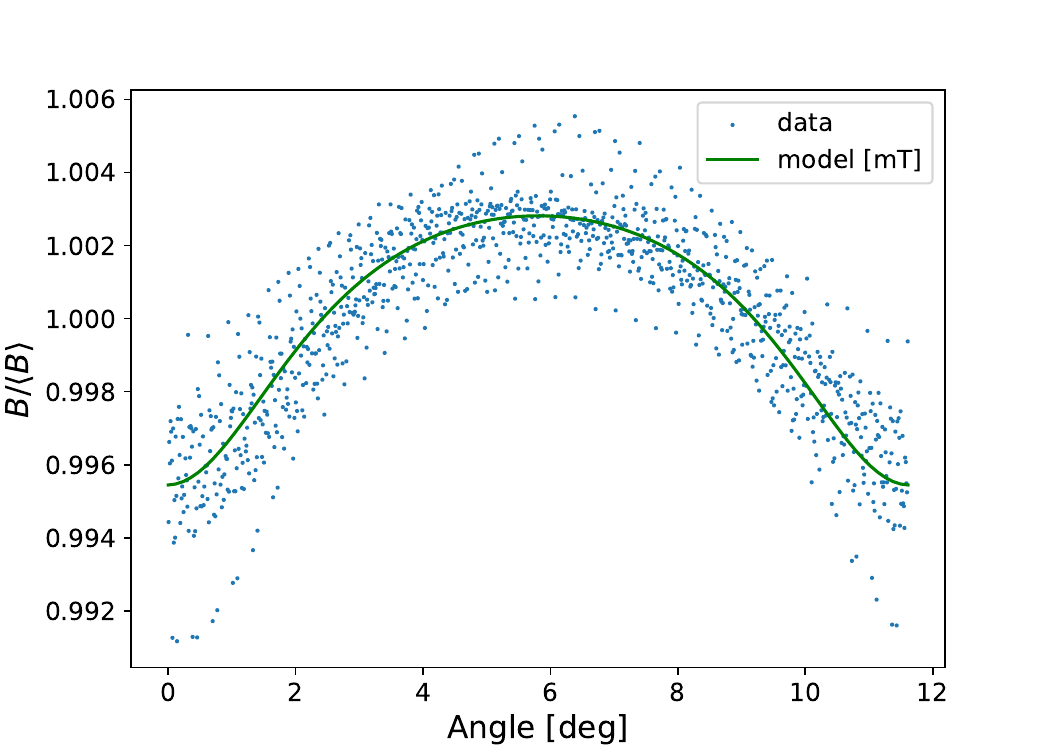}
  \caption{Scatter plot of the magnetic field in the lower panel of Fig.\,\ref{fig:mag_data}, normalized by the overall average. Data from all 31 segments are overplotted into a single plot, and the horizontal axis represents the position within each segment in terms of the rotational angle.The green curve is the normalized magnetic field calculated from Eq.\,\eqref{eq:phi_m}, fit by varying only the gap parameter $\delta$. The best-fit $\delta = 0.28\,\mathrm{mm}$ matches the measured data well.}
  \label{fig:mag_field_model}
\end{figure}

Figure\,\ref{fig:mag_field_model} overplots the magnetic field variation of each segment, where the vertical axis is the measured field $B$ normalized by the overall average magnetic field $\langle B\rangle$, and the horizontal axis represents the position within each segment in terms of the rotational angle.
The green curve in Fig.\,\ref{fig:mag_field_model} is computed by numerically integrating the magnetic scalar potential from Eq.\,\eqref{eq:phi_m} with the gap model.
A value of $\delta = 0.28\,\mathrm{mm}$ reproduces the measured data well.
Although the approximations in Eq.\,\eqref{eq:B_no_gap} and subsequent equations are not used due to the different Hall sensor position and direction, this result confirms that the primary cause of magnetic field non-uniformity is indeed the gap between segments.

\bibliographystyle{aipnum4-1}
\bibliography{MAIN}

\end{document}